\documentstyle[12pt]{article}
\begin{document}
\def\beq{\begin{equation}}
\def\eeq{\end{equation}}
\def\bea{\begin{eqnarray}}
\def\eea{\end{eqnarray}}
\def\ve{\vert}
\def\vel{\left|}
\def\ver{\right|}
\def\nnb{\nonumber}
\def\ga{\left(}
\def\dr{\right)}
\def\aga{\left\{}
\def\adr{\right\}}
\def\rar{\rightarrow}
\def\nnb{\nonumber}
\def\la{\langle}
\def\ra{\rangle}
\def\lla{\left<}
\def\rra{\right>}
\def\ba{\begin{array}}
\def\ea{\end{array}}
\def\tep{$B \rar K \ell^+ \ell^-$}
\def\tepm{$B \rar K \mu^+ \mu^-$}
\def\tept{$B \rar K \tau^+ \tau^-$}
\def\ds{\displaystyle}


\def\lesssim{\mathrel{\mathpalette\vereq<}}
\def\vereq#1#2{\lower3pt\vbox{\baselineskip1.5pt \lineskip1.5pt
\ialign{$\m@th#1\hfill##\hfil$\crcr#2\crcr\sim\crcr}}}

\def\gtrsim{\mathrel{\mathpalette\vereq>}}

\def\alt{\lesssim}
\def\agt{\gtrsim}



\def\bos{\lower 0.5cm\hbox{{\vrule width 0pt height 1.2cm}}}
\def\boss{\lower 0.35cm\hbox{{\vrule width 0pt height 1.cm}}}
\def\aaa{\lower 0.cm\hbox{{\vrule width 0pt height .7cm}}}
\def\dol{\lower 0.4cm\hbox{{\vrule width 0pt height .5cm}}}



\renewcommand{\textfraction}{0.2}    
\renewcommand{\topfraction}{0.8}   
\renewcommand{\bottomfraction}{0.4}   
\renewcommand{\floatpagefraction}{0.8}
\newcommand\mysection{\setcounter{equation}{0}\section}

\def\baeq{\begin{appeq}}     \def\eaeq{\end{appeq}}  
\def\baeeq{\begin{appeeq}}   \def\eaeeq{\end{appeeq}}
\newenvironment{appeq}{\beq}{\eeq}   
\newenvironment{appeeq}{\beeq}{\eeeq}
\def\bAPP#1#2{
 \markright{APPENDIX #1}
 \addcontentsline{toc}{section}{Appendix #1: #2}
 \medskip
 \medskip
 \begin{center}      {\bf\LARGE Appendix  :}{\quad\Large\bf #2}
\end{center}
 \renewcommand{\thesection}{#1.\arabic{section}}
\setcounter{equation}{0}
        \renewcommand{\thehran}{#1.\arabic{hran}}
\renewenvironment{appeq}
  {  \renewcommand{\theequation}{#1.\arabic{equation}}
     \beq
  }{\eeq}
\renewenvironment{appeeq}
  {  \renewcommand{\theequation}{#1.\arabic{equation}}
     \beeq
  }{\eeeq}
\nopagebreak \noindent}

\def\eAPP{\renewcommand{\thehran}{\thesection.\arabic{hran}}}

\renewcommand{\theequation}{\arabic{equation}}
\newcounter{hran}
\renewcommand{\thehran}{\thesection.\arabic{hran}}

\def\bmini{\setcounter{hran}{\value{equation}}
\refstepcounter{hran}\setcounter{equation}{0}
\renewcommand{\theequation}{\thehran\alph{equation}}\begin{eqnarray}}
\def\bminiG#1{\setcounter{hran}{\value{equation}}
\refstepcounter{hran}\setcounter{equation}{-1}
\renewcommand{\theequation}{\thehran\alph{equation}}
\refstepcounter{equation}\label{#1}\begin{eqnarray}}


\newskip\humongous \humongous=0pt plus 1000pt minus 1000pt
\def\caja{\mathsurround=0pt}
\def\eqalign#1{\,\vcenter{\openup1\jot
\caja   \ialign{\strut \hfil$\displaystyle{##}$&$
\displaystyle{{}##}$\hfil\crcr#1\crcr}}\,}


\title{ {\Large {\bf 
More about on the short distance contribution to the 
$B_c \rar B_u^\ast \gamma$ decay } } }

\author{\vspace{1cm}\\
{\small T. M. Aliev \thanks
{e-mail: taliev@metu.edu.tr}\,\,,
M. Savc{\i} \thanks
{e-mail: savci@metu.edu.tr}} \\
{\small Physics Department, Middle East Technical University} \\
{\small 06531 Ankara, Turkey} }
\date{}

\begin{titlepage}
\maketitle
\thispagestyle{empty}

\begin{abstract}
\baselineskip  0.7cm
We calculate the transition form factor for the $B_c \rar B_u^\ast \gamma$
decay taking into account only the short distance contribution, in
framework of QCD sum rules method. We observe that the transition form
factor predicted by the QCD sum rules method is approximately two times 
larger compared to the result predicted by the Isgur, Scora, Grinstein 
and Wise model.  

\end{abstract}
\end{titlepage}

\section{Introduction}
Flavor changing neutral current (FCNC) transitions constitute one of the most
important research area in particle physics. In standard model (SM) these
transitions take place only at one loop level. Therefore the study of the
rare decays allows us to check the gauge structure of the SM and can provide
valuable information for a more precise determination of the
Cabibbo--Kobayashi--Maskawa matrix elements, leptonic decay constants, etc.,
which are poorly known today.

In the SM the FCNC transitions of the down--quark sector have relatively
large branching ratio, due to the large mass of the top quark running in 
the loop and $b \rar s$ transition has already been observed in experiments 
\cite{R1}.
On the other hand, in up quark--sector of the SM these transitions are 
quite rare since in the loop down quark runs
which  has smaller mass compared to top quark mass. 
At present only the upper experimental bounds on the 
FCNC transitions of the up quark sector exist \cite{R2}.

To probe the very rare $c \rar u \gamma$ transition in SM, it was shown in
\cite{R3} that the radiative
beauty conserving $B_c \rar B_u^\ast \gamma$ decay is very promising.
It should be noted that $B_c$ meson has been observed by the CDF
Collaboration at Fermilab \cite{R4}.
This decay receives short and long distance contributions. The short
distance contribution in $B_c \rar B_u^\ast \gamma$ decay comes from FCNC 
$c \rar u \gamma$ transition when $\bar b$ is a spectator quark. The long
distance contributions to $B_c \rar B_u^\ast \gamma$ decay can be grouped
into two classes:

I) Vector meson dominance contribution which corresponds the processes
$c \rar u q_i \bar q_i$, is followed by the conversion 
of $q_i \bar q_i$ pair to
photon while $\bar b$ is spectator again, which is similar to the short
distance contribution case.

II) Annihilation contribution mechanism, which corresponds to the
annihilation process $c\bar b \rar u \bar b$ where photon is attached to any
quark line.

The short and long distance effects to this decay are calculated in framework
of the Isgur, Scora, Grinstein and Wise (ISGW) model \cite{R5} and it is
found that both contributions are comparable to each other which allows, in
principle, probing $c \rar u \gamma$ transition. 
It is found that the branching ratio is of the order of $\sim 10^{-8}$ and
can be detectable at future LHC. This result is quite interesting
and is the first example where short and long distance effects for the 
$c \rar u \gamma$ transition are comparable, contrary to the corresponding
$D$ meson decays for which long distance contribution is dominant
\cite{R6}--\cite{R8}. Therefore this observation opens the way to extract the
short distance $c \rar u \gamma$ contribution from the experiment.
For this reason, in order to
check this principal result it is necessary to perform these calculations
still in another framework. 

In the present letter, we calculate the form factor for the 
$B_c \rar B_u^\ast \gamma$ decay due to the short distance contribution only, 
in frame work of the QCD sum rules. It is observed that the value of the form
factor calculated in the QCD sum rules is approximately two times larger 
compared to the one predicted by the ISGW model. As a result, it seems that in the 
$B_c \rar B_u^\ast \gamma$ decay case, the short and long distance
contributions are of the same order. This circumstance opens the way for a real
possibility of probing rare $c \rar u \gamma$ transition via
$B_c \rar B_u^\ast \gamma$ decay.

As has been noted already, we restrict ourselves only to the short distance
contribution  to the $B_c \rar B_u^\ast \gamma$ decay. The short distance
contribution to the $B_c \rar B_u^\ast \gamma$ decay is obtained from 
$c \rar u \gamma$ transition, where $\bar b$ quark is a spectator. The effective
Hamiltonian for the $c \rar u \gamma$ transition is given as
\bea
{\cal H}_{eff} = - \frac{G_F}{\sqrt{2}} \frac{e}{4 \pi^2} V_{cs} V_{us}^\ast
C_7(\mu) \bar u \sigma_{\mu\nu} \left[ m_c \frac{1+\gamma_5}{2} +
m_u \frac{1-\gamma_5}{2} \right] c \,{\cal F}^{\mu\nu}~,
\eea
where $V_{ij}$ correspond to the CKM matrix elements, and ${\cal
F}_{\mu\nu}$ is the electromagnetic field strength tensor. The appropriate scale
for $C_7(\mu)$ is $\mu=m_c$ since $\bar b$ quark is the spectator 
for the short distance contribution in the $B_c \rar B_u^\ast \gamma$ decay.
In further calculations we will take the mass of the up quark to be zero.

The two
loop QCD corrections to the $c \rar u \gamma$ transition was calculated in 
\cite{R9} whose prediction is $C_7(m_c) = - 0.0068 - 0.02 i$ and this result is
scheme independent.
In order to calculate the amplitude for the $B_c \rar B_u^\ast \gamma$
decay, the matrix elements
\bea
\lla B_u^\ast \vel \bar u \sigma_{\mu\nu} (1\pm\gamma_5)q^\nu \ver B_c \rra~, \nnb 
\eea
need to be calculated at $q^2=0$, where $q$ is the photon four--momentum.
These matrix elements can be written in terms of two gauge invariant form
factors $F_1(0)$ and $F_2(0)$ as follows:
\bea
\lla B_u^\ast(p^\prime,\varepsilon^\prime) \vel \bar u i \sigma_{\mu\nu}
q^\nu c \ver B_c(p) \rra &=& i \epsilon_{\mu\alpha\beta\rho} 
\varepsilon^{\prime\alpha} p^{\prime\beta} q^\rho F_1(0) ~, \nnb \\ \nnb \\
\lla B_u^\ast(p^\prime,\varepsilon^\prime) \vel \bar u i \sigma_{\mu\nu}
q^\nu \gamma_5 c \ver B_c(p) \rra &=& 
\left[ \ga m_{B_c}^2 - m_{B_u^\ast}^2 \dr \varepsilon_\mu^\prime -
\ga \varepsilon^\prime q \dr \ga p + p^\prime \dr_\mu \right] F_2(0) ~.
\eea
Using the relation 
\bea
\sigma_{\mu\nu} \gamma_5 = -\frac{i}{2} \epsilon_{\mu\nu\alpha\beta} 
\sigma^{\alpha\beta} ~,
\eea
one can easily show that $F_2(0)=F_1(0)/2$. Therefore in order to calculate
the short distance part of the $B_c \rar B_u^\ast \gamma$ decay it is enough
to calculate $F_1(0)$ or $F_2(0)$, for which we will employ the three--point
QCD sum rules \cite{R10,R11}.    
For the evolution of the form factor $F_1(0)$ in framework of the QCD sum
rules, we consider the following three--point function 
\bea
\Pi_{\mu\alpha} = - \int d^4x d^4y e^{i(px-p^\prime y)} 
\lla 0 \vel \mbox{\rm T} \Big\{\bar b(y) \gamma_\alpha u(y) \bar u(0) i \sigma_{\mu\nu}
q^\nu c(0) \bar c(x) i \gamma_5 b(x)\Big\} \ver 0 \rra ~,
\eea
where $\bar b \gamma_\alpha u$ and $\bar c i \gamma_5 b$ are the
interpolating currents for states with the $B_u^\ast$ and $B_c$ mesons,
respectively. The Lorentz structure in the correlator (4) can be written as 
\bea
\Pi_{\mu\nu} = i \epsilon_{\mu\nu\alpha\beta} p^\alpha p^{\prime\beta}\, \Pi~,
\eea
where scalar amplitude $\Pi$ is the function of the kinematical 
invariants, i.e., $\Pi = \Pi(p^2,p^{\prime 2})$.

In accordance with the usual QCD sum rules philosophy, the theoretical part
of the three--point correlator can be calculated by employing the operator
product expansion (OPE) for the T--ordered product of currents in (4). The
values of the heavy quark condensates are related to the vacuum expectation
values of the gluon operators. For example
\bea 
\lla Q \bar Q \rra = - \frac{1}{12 m_Q} \frac{\alpha_s}{\pi} 
\lla G^2 \rra - \frac{1}{360 m_Q^3} \frac{\alpha_s}{\pi}
\lla G^2 \rra \cdots ~,
\eea
where $Q$ is the heavy quark and the heavy quark condensate
contributions are suppressed by inverse of the heavy quark mass. for this
reason we safely omit them in our calculations. 

It should be stressed that the light quark condensate does not give any 
contribution to the
above--mentioned decay after double Borel transformation.
Therefore the only non--perturbative contribution to the 
$B_c \rar B_u^\ast \gamma$ decay comes from gluon condensate.

So, in the lowest order of perturbation theory, the three--point function is
given by the bare quark loop and by gluon condensate contribution. The
contribution to the three--point function from the bare loop can be obtained
using the double dispersion representation in $p^2$ and $p^{\prime 2}$ 
\bea
\Pi^{per}(p^2,p^{\prime 2}) = - \frac{1}{4 \pi^2}
\int \displaystyle{\frac{\rho^{per}(p^2,p^{\prime 2})}
{(s-p^2) (s^\prime-p^{\prime 2})}} ds ds^\prime +
\mbox{\rm sub. terms}~.
\eea
The spectral density $\rho^{per}(p^2,p^{\prime 2})$ can be calculated using
the Cutkovsky rule, i.e., by replacing propagators with delta functions:
\bea
\ga k^2 - m_i^2 \dr^{-1} \rar - 2 \pi i \delta(k^2-m_i^2)~. \nnb \\
\eea
After standard calculations for the spectral density we get 
\bea
\rho^{per}(s,s^\prime) = 4 N_c \Big\{ m_c m_b \left[ A_1 + A_2 + {\cal I}_0
\right] - m_c^2 A_1 - 2 A \Big\}~,
\eea
where
\bea
A_1 &=& \frac{2 {\cal I}_0}{(s-s^\prime)^2} \Bigg[ s^\prime
( s+m_b^2-m_c^2 ) - 
\frac{1}{2} (s+s^\prime) (s^\prime+m_b^2) \Bigg]~,\nnb \\ \nnb \\
A_2 &=& \frac{2 {\cal I}_0}{(s-s^\prime)^2} \Bigg[\frac{1}{2}
(s+s^\prime) ( m_c^2 - m_b^2 -s ) + s (s^\prime+m_b^2) \Bigg]~,\nnb \\ \nnb \\
A &=& {\cal I}_0 \,\,\frac{m_c^2 \left[ m_c^2 s^\prime + (m_b^2-s^\prime)
(s-s^\prime)\right]}{2 (s-s^\prime)^2} ~,\nnb \\ \nnb \\
{\cal I}_0 &=& - \frac{1}{4 (s-s^\prime)}~,
\eea
and $N_c$ is the color number.

The region of integration over $s$ and $s^\prime$ is determined by the
following inequalities
\bea
m_b^2 \le s^\prime \le s_0^\prime ~, \nnb \\ \nnb \\
s^\prime - \frac{s^\prime m_c^2}{m_b^2-s^\prime}\le s \le s_0~.
\eea
Note that we have neglected ${\cal O}(\alpha_s/\pi)$ hard gluon corrections to
the triangle diagram, as they are not available yet. However, we expect
their contribution to be about $\sim 10\%$, so that if the accuracy of
the QCD sum rules is taken into account, these corrections would not change
the results drastically.

From our result on spectral density we can get the spectral density for 
$B \rar K^\ast \gamma$ decay (when $m_s \rar 0$) if we formally make the 
replacements $m_b \rar 0$ and $m_c \rar m_b$ in Eqs. (9) and (10). 
Indeed, our results coincide with the results of \cite{R12} for 
$B \rar K^\ast \gamma$ decay, after the above--mentioned substitutions are
performed.
The gluon condensate contribution to three--point correlator
(4) is given by diagrams depicted in Fig. (1). The calculations of these
diagrams were carried out in the Fock--Schwinger fixed point gauge 
\cite{R13,R14,R15}; $x^\mu A_\mu (x) = 0$. 
For calculation of the gluon condensate contributions, we have used the
Schwinger representation for the Euclidean propagators, i.e.,
\bea
\frac{1}{\Big[ k^2 + m^2 \Big]^a} = \frac{1}{\Gamma(a)} \int_0^\infty
d\alpha \, \alpha^{n-1} e^{-\alpha (k^2+m^2)}~,
\eea
which is very suitable for applying the Borel transformation, since
\bea
\hat B_{p^2}(M^2)\, e^{-\alpha p^2} = \delta(1-\alpha M^2)~.
\eea
The analytical expression for the Wilson coefficient of the gluon condensate
operator $C_{G^2}$ is quite lengthy and for this reason it is  presented in the
appendix. 

It should be noted that Borel transformed Wilson coefficient of the gluon
condensate contribution in the three point sum rules with arbitrary mass,
which appears in the study of the form factors for the vector and axial
vector current transitions of the semileptonic $B_c \rar J/\psi \ell \nu$
decay, was investigated in detail in \cite{R16}. 

We now turn our attention to the computation of the physical part of the sum
rules. Assuming that the spectral density is well convergent, the physical
spectral density is saturated by the lowest lying hadronic states
plus a continuum starting at some effective thresholds $s_0,~s_0^\prime$
\bea
\rho^{phy}_{\mu\alpha}(s,s^\prime) = \rho^{res}_{\mu\alpha}(s,s^\prime) + 
\theta(s-s_0) \theta(s^\prime-s_0^\prime) \rho^{cont}_{\mu\alpha}(s,s^\prime)~,
\eea
where 
\bea
\rho^{res}_{\mu\alpha} &=& \lla 0 \vel \bar c i \gamma_5 b \ver B_c \rra 
\lla B_c \vel \bar u i \sigma_{\mu\nu} q^\nu c \ver B_u^\ast \rra 
\lla B_u^\ast \vel \bar b \gamma_\alpha u \ver 0 \rra \nnb \\
&\times&(2 \pi)^2 \delta (s - M_{B_c}^2) \delta (s^\prime - M_{B_u^\ast}^2)~,
\eea
and $\rho^{cont}$ corresponds to the continuum contribution.
The matrix elements in (15) are defined in the following way:
\bea
\lla 0 \vel \bar c i \gamma_5 b \ver B_c \rra &=& 
\frac{f_{B_c} m_{B_c}^2}{m_b+m_c} ~, \nnb \\
\lla B_u^\ast \vel \bar b \gamma_\alpha u \ver 0 \rra &=&
f_{B_u^\ast} m_{B_u^\ast} \varepsilon_\alpha^\ast~. \nnb 
\eea
Selecting the structure 
$i \epsilon_{\mu\nu\alpha\beta} p^\alpha p^{\prime \beta}$ for 
$\rho^{res}$, we have
\bea
\rho^{res} = \frac{f_{B_c} m_{B_c}^2}{m_b+m_c} f_{B_u^\ast} m_{B_u^\ast}
F_1(0) (2 \pi)^2 \delta(s-M_{B_c}^2) \delta (s^\prime - M_{B_u^\ast}^2)~. \nnb
\eea
The continuum contribution is modeled as a perturbative contribution
starting from thresholds $s_0$ and $s_0^\prime$. Equating (15) to the
theoretical part contribution (7) and performing double Borel transformations
with respect to the parameters $p^2$ and $p^{\prime 2}$, we finally get the
following sum rule for the transition form factor:
\bea
F_1(0) &=& \frac{(m_b+m_c)}{f_{B_u^\ast} m_{B_u^\ast} f_{B_c} m_{B_c}^2}
e^{m_{B_c}^2/M_1^2} e^{m_{B_u^\ast}^2/M_2^2} \nnb \\
&\times& \Bigg\{
- \frac{1}{4\pi^2} \int ds ds^\prime \rho(s,s^\prime) 
e^{-s/M_1^2} e^{-s^\prime/M_2^2} +
M_1^2 M_2^2 \la \frac{\alpha_s}{\pi} G^2 \ra 
C_{G^2} \Bigg\}~,
\eea
where $C_{G^2}$ is the Wilson coefficient of the gluon condensate contribution.
This expression is the final result for the transition form factor evaluated
at $q^2=0$. 

For the numerical
analysis we have used the following values of the input parameters that
enter into sum rules (16): $f_{B_c}=385~MeV$ \cite{R17}, 
$f_{B_u^\ast}=160~MeV$ \cite{R18}, $m_c=1.4~GeV$,  
$\la \alpha_s G^2/\pi\ra=0.012~GeV^4$ \cite{R10}, $s_0=50~GeV^2$ and
$s_0^\prime=35~GeV^2$.
As is obvious, Eq. (16) involves two independent Borel
parameters $M_1^2,~M_2^2$, and then the main problem is finding the region
where the dependence of these parameters is weak and at the same time power
corrections and the continuum remains under control. 

In Fig. (2) we present the dependence of the transition form factor $F_1(0)$
on $M_1^2$ and $M_2^2$. Numerical calculations show that the best stability
for the form factor $F_1(0)$ is achieved for 
$15~GeV^2 \le M_1^2 \le 20~GeV^2$ and $8~GeV^2 \le M_2^2 \le 12~GeV^2$. Our
final result for the form factor is
\bea
F_1(0) = (0.9 \pm 0.1)~. \nnb
\eea
For a comparison, we note that the IGSW model's prediction for this
transition form factor is $F_1(0) = 0.48$ \cite{R3}. Therefore the 
branching ratio in our case is approximately four times larger compared to that
predicted in \cite{R3}. It should be stressed again that in the present
work only the short distance contribution to the 
$B_c \rar B_u^\ast \gamma$ decay is considered.      

Using this result we observe that the short distance contribution to the
branching ratio is of the order of $\sim 2\times10^{-8}$, which can be quite
detectable at future LHC. Moreover our result show that the
short distance contribution in our approach is comparable or larger than the
long distance contribution calculated in \cite{R3}
\bea
{\cal B} \ga B_c \rar B_u^\ast \gamma \dr = 
\ga 7.5^{\displaystyle{+7.7}}_{\displaystyle{-4.3}} \dr \times
10^{-9}, \nnb
\eea  
and our result proves that there is indeed real possibility
for probing the $c \rar u \gamma$ decay via the beauty conserving 
$B_c \rar B_u^\ast \gamma$ decay.

As the final remark we would like to note that the approach presented in
this work is applicable for calculating the short distance contributions to
the branching ratio of the $B_s \rar B_d^\ast \gamma$ and
$B_c \rar D_s^\ast \gamma$ decays.
 
\newpage
\bAPP{A}{The gluon condensate contribution}

In this section we will present the explicit expressions for the Wilson
coefficients $C_{G^2}$ for each diagram, after
the Borel transformation with respect to $p^2$ and $p^{\prime 2}$, which are
presented in Fig. (1). 

\baeq\eqalign{
\lefteqn{
( C_{G^2} )_1 =
96 m_c \Big\{\Big[m_b \Big(I_0[1, 3, 1] + m_c^2 I_0[1, 4, 1] + I_1[1, 3, 1] 
+ m_c^2 I_1[1, 4, 1] } \cr  
&+ I_2[1, 3, 1] 
- m_c^2 I_2[1, 4, 1]\Big)\Big] + 
    m_c \Big(I_1[1, 3, 1] + m_c^2 I_1[1, 4, 1] 
+ 2 I_3[1, 4, 1]\Big)\Big\}~, 
}\eaeq
\\
\baeq\eqalign{
\lefteqn{
( C_{G^2} )_2 =
16 \Big\{2 I_0[1, 1, 2] + 2 m_c m_b I_0[1, 1, 3] + 2 I_0[2, 1, 1] + 
    3 m_c m_b I_0[2, 1, 2]} \cr 
&+ 4 m_b^2 I_0[2, 1, 2] + 4 m_c m_b^3 I_0[2, 1, 3] + 
    2 m_c m_b I_0[3, 1, 1] + 2 m_b^2 I_0[3, 1, 1] \cr 
&+ 6 m_c m_b^3 I_0[3, 1, 2] + 
    2 m_b^4 I_0[3, 1, 2] + 2 m_c m_b^5 I_0[3, 1, 3] + 2 I_1[1, 1, 2] \cr 
&- 2 m_c^2 I_1[1, 1, 3] + 2 m_c m_b I_1[1, 1, 3] + 2 I_1[2, 1, 1] - 
    m_c^2 I_1[2, 1, 2] \cr 
&+ m_c m_b I_1[2, 1, 2] + 4 m_b^2 I_1[2, 1, 2] - 
    4 m_c^2 m_b^2 I_1[2, 1, 3] + 4 m_c m_b^3 I_1[2, 1, 3] \cr 
&- 2 m_c^2 I_1[3, 1, 1] + 
    2 m_b^2 I_1[3, 1, 1] - 6 m_c^2 m_b^2 I_1[3, 1, 2] + 4 m_c m_b^3 I_1[3, 1, 2] \cr 
&+ 2 m_b^4 I_1[3, 1, 2] - 2 m_c^2 m_b^4 I_1[3, 1, 3] + 2 m_c m_b^5 I_1[3, 1, 3] + 
    2 m_c m_b I_2[1, 1, 3] \cr 
&+ I_2[2, 1, 1] + m_c m_b I_2[2, 1, 2] + 
    4 m_c m_b^3 I_2[2, 1, 3] - 4 m_b^2 I_2[3, 1, 1] \cr 
&+ 4 m_c m_b^3 I_2[3, 1, 2] + 
    2 m_c m_b^5 I_2[3, 1, 3] - 4 I_3[1, 1, 3] - 4 I_3[2, 1, 2] \cr 
&- 8 m_b^2 I_3[2, 1, 3] - 8 I_3[3, 1, 1] - 16 m_b^2 I_3[3, 1, 2] - 
    4 m_b^4 I_3[3, 1, 3]\Big\} \cr 
&- 32 M_2^2\frac{d}{dM_2^2} \Big\{M_2^2 \Big[I_0[2, 1, 2] + 2 m_c m_b I_0[2, 1, 3] + 
       2 m_c m_b I_0[3, 1, 2] \cr 
&+ m_b^2 I_0[3, 1, 2]  
+ 2 m_c m_b^3 I_0[3, 1, 3] + 
       I_1[2, 1, 2] - 2 m_c^2 I_1[2, 1, 3] + 2 m_c m_b I_1[2, 1, 3] \cr 
&- 2 m_c^2 I_1[3, 1, 2] + m_c m_b I_1[3, 1, 2] + m_b^2 I_1[3, 1, 2] - 
       2 m_c^2 m_b^2 I_1[3, 1, 3] \cr 
&+ 2 m_c m_b^3 I_1[3, 1, 3] - I_2[2, 1, 2] + 
       2 m_c m_b I_2[2, 1, 3] - 2 I_2[3, 1, 1] \cr 
&+ m_c m_b I_2[3, 1, 2] - 
       m_b^2 I_2[3, 1, 2] + 2 m_c m_b^3 I_2[3, 1, 3] - 4 I_3[2, 1, 3] \cr 
&- 6 I_3[3, 1, 2] - 4 m_b^2 I_3[3, 1, 3]\Big]\Big\} \cr
&- 32 M_2^4 \ga \frac{d^2}{dM_2^2} \dr^2 \Big\{ M_2^4 \Big[ 
     m_c^2 I_1[3, 1, 3] +2 I_2[3, 1, 2]  
- m_c m_b \Big(I_0[3, 1, 3] + I_1[3, 1, 3] \cr 
&+ I_2[3, 1, 3]\Big) + 2 I_3[3, 1, 3]\Big] 
     \Big\}~, 
}\eaeq
\\
\baeq\eqalign{
\lefteqn{    
( C_{G^2} )_3  =
96 m_b \Big\{-\Big( m_c^2 m_b I_1[4, 1, 1]\Big) +
    m_c \Big(I_0[3, 1, 1] + m_b^2 I_0[4, 1, 1] } \cr
&+ I_1[3, 1, 1] + m_b^2 I_1[4, 1, 1] +
       I_2[3, 1, 1] + m_b^2 I_2[4, 1, 1]\Big) - 2 m_b I_3[4, 1, 1]\Big\}~,
}\eaeq
\\
\baeq\eqalign{
\lefteqn{
( C_{G^2} )_4 =
-32 m_c m_b \Big\{ I_0[2, 1, 2] + I_0[3, 1, 1] + m_b^2 I_0[3, 1, 2] +
I_1[2, 1, 2] + m_b^2 I_1[3, 1, 2] } \cr
&+ I_2[2, 1, 2] + m_b^2 I_2[3, 1, 2] - 4 I_3[3, 1, 2]
- M_2^2 \frac{d}{dM^2} \Big[ M_2^2 \Big( I_0[3, 1, 2] + I_1[3, 1, 2] \cr &
+ I_2[3, 1, 2] \Big)\Big]  \Big\}
+16 \Big\{m_c^2 I_1[2, 1, 2]
+ I_2[2, 1, 1]- m_c m_b \Big(I_0[2, 1, 2] + I_1[2, 1, 2]\cr
&+ I_2[2, 1, 2]\Big)
+ 4 I_3[2, 1, 2]\Big\} ~,
}\eaeq
\\
\baeq\eqalign{
\lefteqn{
( C_{G^2} )_5 =
16 \Big\{2 I_0[1, 1, 2] + I_0[1, 2, 1] + 2 m_c^2 I_0[1, 2, 2] + m_c m_b I_0[1, 2, 2] } \cr 
&+ m_b^2 I_0[2, 1, 2] + m_c m_b I_0[2, 2, 1] + m_b^2 I_0[2, 2, 1] + 
    m_c^2 m_b^2 I_0[2, 2, 2] \cr 
&+ m_c m_b^3 I_0[2, 2, 2] + I_1[1, 1, 2] + I_1[1, 2, 1] + 
    m_c^2 I_1[1, 2, 2] \cr 
&+ m_c m_b I_1[1, 2, 2] - I_1[2, 1, 1] - 
    2 m_c m_b I_1[2, 1, 2] - m_c^2 I_1[2, 2, 1] \cr 
&+ m_b^2 I_1[2, 2, 1] - 
    2 m_c^3 m_b I_1[2, 2, 2] + m_c m_b^3 I_1[2, 2, 2] + I_2[1, 1, 2] \cr 
&+ I_2[1, 2, 1] + 
    m_c^2 I_2[1, 2, 2] + m_c m_b I_2[1, 2, 2] - 2 m_c m_b I_2[2, 2, 1] \cr 
&+ m_b^2 I_2[2, 2, 1] + m_c m_b^3 I_2[2, 2, 2] - 2 I_3[2, 1, 2] - 4 I_3[2, 2, 1] \cr 
&- 2 m_c^2 I_3[2, 2, 2] - 4 m_c m_b I_3[2, 2, 2] \Big\} \cr 
&- 16 M_1^2\frac{d}{dM_1^2} \Big\{ M_1^2 \Big[
    I_2[2, 2, 1] + 2 I_3[2, 2, 2]\Big]\Big\} \cr 
&+ 16 M_2^2 \frac{d}{dM_2^2} \Big\{ M_2^2\Big[-m_c m_b I_0[2, 2, 2] + I_1[2, 1, 2] 
+ m_c^2 I_1[2, 2, 2] - m_c m_b I_1[2, 2, 2] \cr 
&+ I_2[2, 1, 2] + I_2[2, 2, 1] + m_c^2 I_2[2, 2, 2] - 
       m_c m_b I_2[2, 2, 2] + 2 I_3[2, 2, 2] \Big] \Big\}\cr 
&+ 16 \Big\{I_1[1, 1, 2] + m_c^2 I_1[1, 2, 2] 
+ I_2[1, 2, 1] - m_c m_b \Big(I_0[1, 2, 2] + I_1[1, 2, 2] \cr 
&+ I_2[1, 2, 2]\Big)
+ 2 I_3[1, 2, 2]\Big\}~,
}\eaeq
\\
\baeq\eqalign{
\lefteqn{
( C_{G^2} )_6  =
16 \Big\{2 I_0[1, 2, 1] + 2 I_0[2, 1, 1] + 2 m_c^2 I_0[2, 2, 1] } \cr  
&+ 6 m_c m_b I_0[2, 2, 1]  
+ 2 m_b^2 I_0[2, 2, 1] + 2 I_1[1, 2, 1] - 
    5 I_1[2, 1, 1] - 5 m_c^2 I_1[2, 2, 1] \cr 
&+ 6 m_c m_b I_1[2, 2, 1] + 
    2 m_b^2 I_1[2, 2, 1] + 2 I_2[1, 2, 1] - I_2[2, 1, 1] \cr 
&- m_c^2 I_2[2, 2, 1] + 
    6 m_c m_b I_2[2, 2, 1] + 2 m_b^2 I_2[2, 2, 1] - 14 I_3[2, 2, 1] \Big\}\cr 
&- 32 M_1^2\frac{d}{dM_1^2} \Big\{ M_1^2 \Big[I_0[2, 2, 1] + I_1[2, 2, 1] + 
I_2[2, 2, 1] \Big]\Big\}~,
}\eaeq
where the subscripts in the Wilson coefficients $C_{G^2}$ 
denote the corresponding diagrams in Fig. (1). 
In calculating the gluon condensate contribution, we need integrals of the
following types:

\baeq\eqalign{
I_0[a,b,c] &= 
\int \frac{d^4k}{(2 \pi)^4} \frac{1}{\left[ k^2-m_b^2 \right]^a
\left[ (p+k)^2-m_c^2 \right]^b \left[ (p^\prime+k)^2\right]^c}~,
}\eaeq
\\
\baeq\eqalign{
I_\mu[a,b,c] &= 
\int \frac{d^4k}{(2 \pi)^4} \frac{k_\mu}{\left[ k^2-m_b^2 \right]^a  
\left[ (p+k)^2-m_c^2 \right]^b \left[ (p^\prime+k)^2\right]^c}~,
}\eaeq
\\
\baeq\eqalign{
I_{\mu\nu}[a,b,c] &= 
\int \frac{d^4k}{(2 \pi)^4} \frac{k_\mu k_\nu}{\left[ k^2-m_b^2 \right]^a
\left[ (p+k)^2-m_c^2 \right]^b \left[ (p^\prime+k)^2\right]^c}~.
}\eaeq
The integrals $I_\mu$ and $I_{\mu\nu}$ can be written in the following form
\baeq\eqalign{
I_\mu &= I_1 p_\mu + I_2 p_\mu^\prime~, \cr
I_{\mu\nu} &= I_3 g_{\mu\nu} + I_4 p_\mu p_\nu + 
I_5 p_\mu^\prime p_\nu^\prime + I_6 p_\mu p_\nu^\prime+
I_7 p_\mu^\prime p_\nu~.
}\eaeq
It should be noted that only the $g_{\mu\nu}$ term in $I_{\mu\nu}$ gives
contribution to  the $\epsilon_{\mu\nu\alpha\beta}$ structure which we need
in our analysis.

After double Borel transformations with respect to the variables $p^2$ and 
$p^{\prime 2}$, the explicit forms of the integrals $I_0[a,b,c]$, $I_1[a,b,c]$,
$I_2[a,b,c]$ and $I_3[a,b,c]$ are as follows (see also \cite{R16})
\baeq\eqalign{
I_0[a,b,c] &= \frac{(-1)^{a+b+c}}{16 \pi^2\,\Gamma(a) \Gamma(b) \Gamma(c)}
(M_1^2)^{2-a-b} (M_2^2)^{2-a-c} \, {\cal U}_0(a+b+c-4,1-c-b)~,
}\eaeq
\\
\baeq\eqalign{
I_1[a,b,c] &= \frac{(-1)^{a+b+c+1}}{16 \pi^2\,\Gamma(a) \Gamma(b) \Gamma(c)}
(M_1^2)^{2-a-b} (M_2^2)^{3-a-c} \, {\cal U}_0(a+b+c-5,1-c-b)~,                
}\eaeq
\\
\baeq\eqalign{
I_2[a,b,c] &= \frac{(-1)^{a+b+c+1}}{16 \pi^2\,\Gamma(a) \Gamma(b) \Gamma(c)}
(M_1^2)^{3-a-b} (M_2^2)^{2-a-c} \, {\cal U}_0(a+b+c-5,1-c-b)~,
}\eaeq
\\
\baeq\eqalign{
I_3[a,b,c] &= \frac{(-1)^{a+b+c+1}}{32 \pi^2\,\Gamma(a) \Gamma(b) \Gamma(c)}
(M_1^2)^{3-a-b} (M_2^2)^{3-a-c} \, {\cal U}_0(a+b+c-6,2-c-b)~.
}\eaeq
The function ${\cal U}_0(\alpha,\beta)$ is given by the following
expression
\baeq\eqalign{
{\cal U}_0(\alpha,\beta) = \int_0^\infty dy (y+M_1^2+M_2^2)^\alpha y^\beta
\,exp\left[ -\frac{B_{-1}}{y} - B_0 - B_1 y \right]~,
}\eaeq
where 
\baeq\eqalign{
B_{-1} &= \frac{m_c^2}{M_1^2} \left[M_1^2 + M_2^2 \right] ~, \cr
B_0 &= \frac{1}{M_1^2 M_2^2} \left[M_1^2  m_b^2 + M_2^2 (m_b^2+m_c^2)
\right] ~, \cr
B_1 &=\frac{m_b^2}{M_1^2 M_2^2}~.
}\eaeq

\eAPP

\newpage
\section*{Figure captions}
{\bf Fig. 1} Gluon condensate contribution diagrams to the 
$B_c \rar B_u^\ast \gamma$ decay. In this figure the dashed line represents
the soft gluon line, $c,~u,~b$ identify the quark lines, $p$ and
$p^\prime$ are the four--momenta of the incoming $B_c$ and outgoing 
$B_u^\ast$ mesons, respectively, and $q$ is the four--momentum of the
outgoing photon.  \\ \\
{\bf Fig. 2} The dependence of the transition form factor $F_1(0)$ on the Borel
parameters $M_1^2$ and $M_2^2$.

\newpage
\begin{figure}[H]
\vskip 1.7 cm
    \includegraphics{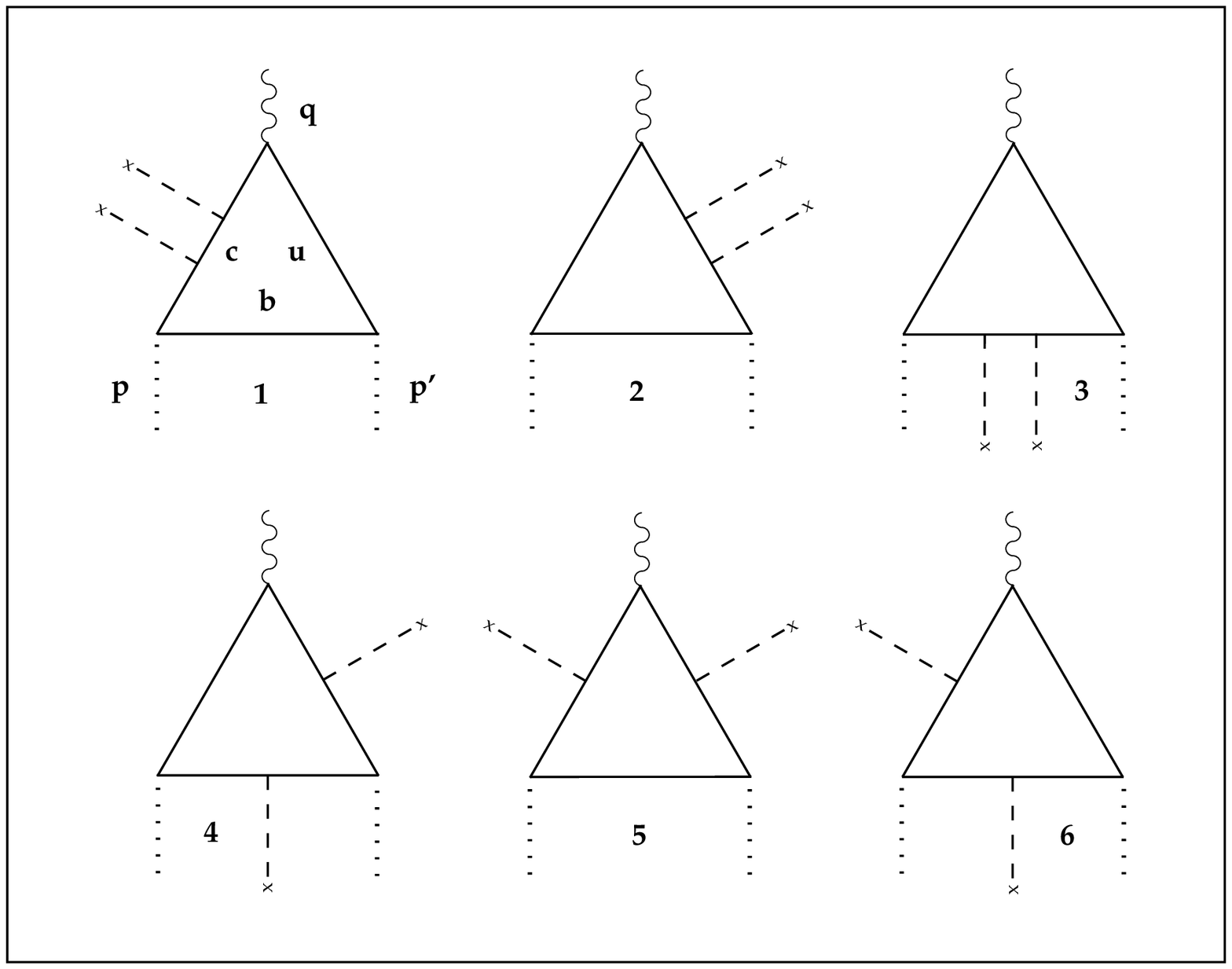}
\vskip 9cm   
\caption{}
\end{figure}

\begin{figure}      
\vskip -0.5cm
       \includegraphics{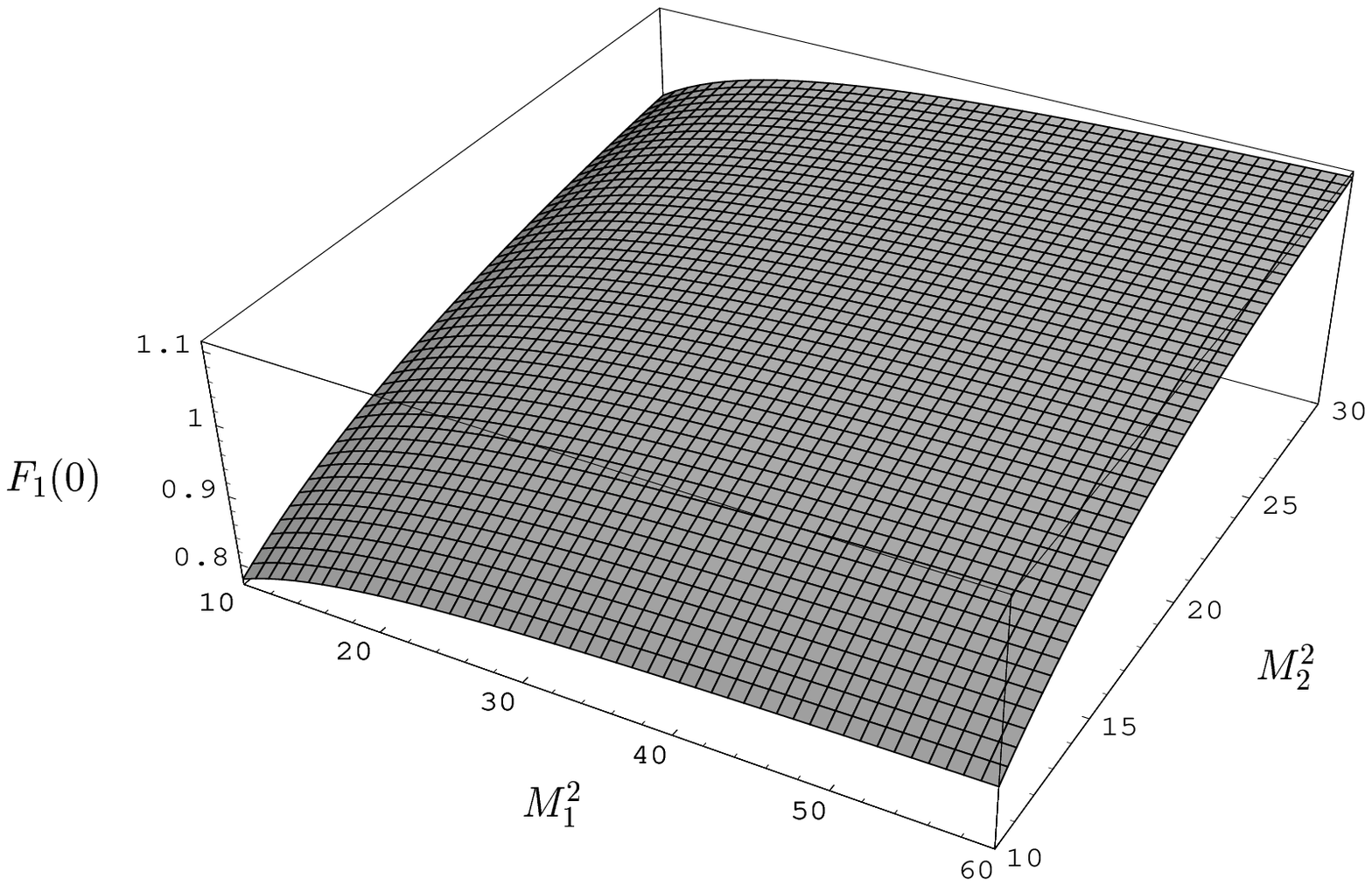}
\vskip 9.5 cm
\caption{ }
\end{figure}

\end{document}